\documentclass[a4paper,10pt]{article}
\usepackage{url}
\usepackage{times,epsfig,amsfonts,amsmath,dcolumn,enumerate,multirow,graphicx,amssymb,colortbl,bm,url}
\usepackage{mathtools,upgreek}
\usepackage{mathtools}
\usepackage{afterpage}
\usepackage{natbib}
\usepackage[margin=1in]{geometry}

\DeclareMathAlphabet{\pazocal}{OMS}{zplm}{m}{n}

\newcommand{\Lb}{\pazocal{L}}

\newcommand{\bx}{{\bf x}}

\newcommand{\bz}{{\bf z}}

\begin{document}

\title{On models for the estimation of the excess mortality hazard in case of insufficiently stratified life tables}

%

\author{Francisco J. Rubio$^{a\ast}$, Bernard Rachet$^b$, Roch Giorgi$^c$,  Camille Maringe$^b$, \\
  Aur{\'e}lien Belot$^b$, and the CENSUR working survival group\\
\small $^{a}$  Department of Mathematics,
\\ \small King's College London,
\\ \small London, WC2R 2LS, UK.
\\ \small $^{b}$ Cancer Survival Group, Faculty of Epidemiology and Population Health,
\\ \small Department of Non-Communicable Disease Epidemiology,
\\ \small London School of Hygiene \& Tropical Medicine,
\\ \small London, WC1E 7HT, UK.
\\ \small $^{c}$ Aix Marseille Univ, APHM, INSERM, IRD, SESSTIM,
\\ \small Sciences Economiques \& Sociales de la Sant{\'e} \& Traitement de l'Information M{\'e}dicale,
\\ \small Hop Timone, BioSTIC, Biostatistique et Technologies de l'Information et de la Communication,
\\ \small Marseille, France.\\ 
\small $^{*}$ E-mail: javier.rubio\_alvarez@kcl.ac.uk}
\maketitle

\begin{abstract}
{In cancer epidemiology using population-based data, regression models for the excess mortality hazard is a useful method to estimate cancer survival and to describe the association between prognosis factors and excess mortality. This method requires expected mortality rates from general population life tables: each cancer patient is assigned an expected (background) mortality rate obtained from the life tables, typically at least according to their age and sex, from the population they belong to. However, those life tables may be insufficiently stratified, as some characteristics such as deprivation, ethnicity, and comorbidities, are not available in the life tables for a number of countries. This may affect the background mortality rate allocated to each patient, and it has been shown that not including relevant information for assigning an expected mortality rate to each patient induces a bias in the estimation of the regression parameters of the excess hazard model. We propose two parametric corrections in excess hazard regression models, including a single-parameter or a random effect (frailty), to account for possible mismatches in the life table and thus misspecification of the background mortality rate. In an extensive simulation study, the good statistical performance of the proposed approach is demonstrated, and we illustrate their use on real population-based data of lung cancer patients. We present conditions and limitations of these methods, and provide some recommendations for their use in practice.}

\end{abstract}

%
%
\section{Net survival and excess mortality hazard model}
Survival analysis after the diagnosis of cancer is an active research area in cancer epidemiology and a primary interest of many countries. The three typical frameworks adopted to model cancer survival data are: (i) the overall survival setting (where \emph{overall or all-cause} mortality is studied), (ii) the cause-specific setting (where the cause of death is known), and (iii) the relative survival setting. The overall survival setting is not the optimal choice for cancer epidemiology when the main interest is on comparing two or more populations (\emph{e.g.}~two countries, two periods in the same country, two groups with different deprivation levels within the same country and at the same period, and etcetera), since it is affected by other causes of mortality, which may differ for the populations of interest. In practice, the cause of death is typically unavailable or, in the absence of a standardised protocol, not reliable. Thus, the cause-specific setting may not be a reasonable choice either. The relative survival setting \citep{perme:2016} represents a useful alternative, where the mortality hazard associated to other causes is approximated by the general population hazard $h_P(\cdot;\bz)$, which is typically obtained from life tables based on the available sociodemographic characteristics $\bz$ (\textit{e.g.}~sex, region and deprivation level) in addition to age and year. The general population hazard is also referred to as the ``expected hazard'' and the ``background mortality rate''. While defined in a hypothetical world, where patients could only die from the cancer under study, net survival (the main object of interest in the relative survival setting) represents a useful way of reporting and comparing the probability of survival of cancer patients since this quantity is not affected by differences in expected mortality (due to other causes) between populations. The basic idea behind net survival consists of decomposing the hazard function associated to an individual, $h_o(t;\bx)$, as the sum of the hazard associated to the disease of interest (\textit{e.g.}~a specific cancer), $h_E(t;\bx)$, and the hazard associated to other causes in the population of interest, $h_{Other}(t;\bz)$. The hazard associated to other causes is approximated with the general population hazard $h_P(t;\bz)$, assuming that the contribution in the general population hazard of a specific cancer type is small compared to all other causes of death. This is:
\begin{eqnarray}\label{HazardAddModel}
h_o(t;\bx) = h_P(A+t;y+t;\bz) + h_E(t;\bx),
\end{eqnarray}
where $\bx$ and $\bz$ are vectors of covariates, and $\bz$ typically corresponds to a subset of covariates of $\bx$. The variables ``$A$'' and ``$y$'' represent the age at diagnosis and the year of diagnosis, respectively, thus $A+t$ and $y+t$ represent the age and the year at time $t$ after diagnosis. This model is also known as \emph{excess hazard model} \citep{E90}. The net survival is defined as the survival function associated to the excess hazard function $h_E(\cdot;\bx)$. Estimation of the net survival function has been largely studied from parametric, semiparametric, and nonparametric perspectives (see \citealp{R07}, \citealp{Pohar2009}, \citealp{PS12}, and \citealp{R18}).

In practice, a limitation of quantities derived in the relative survival setting \citep{PS12}, such as the excess hazard, is that patients need to be matched to groups of the general population sharing the available characteristics ${\bf z}$ in order to obtain the background mortality rates $h_P(\cdot;\bz)$ from the life tables. The number of available characteristics ${\bf z}$ varies for different countries, and there exist certain characteristics that are not available at the population level that may affect the background mortality rates such as deprivation, ethnicity, drug use (tobacco, alcohol, and etcetera), comorbidities, among others. For instance, it has been shown that deprivation levels are associated with life expectancy in some populations \citep{W05}. Thus, if life tables are obtained without a proper stratification of deprivation levels, this will imply that two individuals with different deprivation levels (\textit{e.g.}~most affluent \textit{vs.}~most deprived), but sharing other socio-demographic characteristics, will be assigned the same background mortality. We will refer to the case when the life table is not sufficiently stratified as a ``mismatch in the life table''. This mismatch may also apply to other characteristics (\textit{e.g.}~smoking status), thus potentially impacting the comparison ability of the net survival measure \citep{P18}. Moreover, \cite{D98} and \cite{G12} showed that not including relevant information for matching the mortality rates of patients induces a bias in the estimation of the parameters in net survival models (even for other variables than those included in the life tables). Concerns about the implications of this kind of mismatches in the life tables have recently been discussed in \cite{P18} and \cite{bower:2018}. Thus, it is desirable to produce models that can capture possible mismatching information in the background mortality rates associated to each patient, and that allow the assessment of the impact of this mismatch.

In this work, we study an existing model for correcting the background mortality using a single correction parameter \citep{CR91}, and extend it to a general hazard structure for excess hazard regression models \citep{R18}. We also propose a correction model using a random effect (instead of a single parameter). For each of these models, we study the properties of maximum likelihood estimation of the corresponding parameters. We assess the performance of this inferential procedure in an extensive simulation study. We apply and compare these methods using a lung cancer data example. We conclude with some practical advice and summarise the conditions and limitations of these methods, as well as potential directions for further research.

\section{Corrections of the background mortality in excess hazard regression}
In this section, we present two parametric corrections that can account for mismatches in the life table. The first one corresponds to the single parameter correction proposed by \cite{CR91}. The second one corresponds to our proposal, which assumes that the correction to the background mortality is random and can be modelled with a parametric distribution. Next, we present a brief summary of these methods, as well as some of their properties and limitations.

\subsection{Single parameter correction: Cheuvart and Ryan's approach}
In the context of clinical trials, \cite{CR91} proposed an extension of model \eqref{HazardAddModel} in which they allowed for a constant correction on the population hazard: the overall hazard after $t$ years of follow-up for a patient who entered the trial at age $A$ in year $y$ and with covariates ${\bx}$ was assumed to be
\begin{eqnarray}\label{CheuvartHazard}
h_o^{C}(t;\bx) &=& \gamma h_P(A+t;y+t;\bz) + h_E(t;\bx),
\end{eqnarray}
where $\gamma>0$ is an unknown parameter differentiating the competing mortality of eligible patients from that of the general population. \cite{CR91} employ the proportional excess hazard model $h_E(t;\bx) = h_{E,0}(t;\eta)\exp({\bf z}^{\top}\bm{\beta})$, where $h_{E,0}(\cdot;\eta)$ is the baseline excess hazard with parameter $\eta$, $\bm{\beta}$ is a vector of regression parameters, including the effects of treatment and other prognostic factors. In the context of clinical trials, the correction parameter can be reasonably assumed to be the same for all individuals since the population entering the trials is selected based on their characteristics (usually patients without comorbid conditions, and etcetera).

Model \eqref{CheuvartHazard} can be rewritten in terms of the cumulative hazard and the survival functions as follows
\begin{eqnarray}\label{CheuvartCHSurv}
H_o^{C}(t;\bx,\gamma) &=& \gamma [H_P(A+t;y+t;\bz)-H_P(A;y;\bz)] + H_E(t;\bx),\\
S_o^{C}(t;\bx,\gamma) &=& \exp\{-\gamma [H_P(A+t;y+t;\bz)-H_P(A;y;\bz)]\} \exp\left[-H_E(t;\bx)\right],
\end{eqnarray}
where $H_P$ and $H_E$ are the cumulative hazard functions obtained from $h_P$ and $h_E$, respectively.

In principle, model \eqref{CheuvartHazard} can also be used to account for mismatched life tables, as the correction is made on the population hazard. The basic assumption behind this model is that the true competing mortality is proportional to the mortality of the population obtained from the life tables, and the correction is the same for all individuals. Expression \eqref{CheuvartCHSurv} also indicates that the information used for estimating the additional correction parameter $\gamma$ comes from the differences in the cumulative hazard $H_P(A+t;y+t;\bz)-H_P(A;y;\bz)$, which are not used in the estimation of the classical model \eqref{HazardAddModel}.

\subsection{Frailty correction for the population hazard}

One limitation of model \eqref{CheuvartHazard} is that the correction $\gamma$, made on the population hazard, is assumed to be constant across all the individuals. This assumption may not be realistic in population studies, where the diversity of unavailable sociodemographic characteristics used to obtain the life tables may induce a non-constant mismatch. Thus, instead of assuming that $\gamma$ is constant in model \eqref{CheuvartHazard}, we assume that $\gamma$ is a positive continuous random variable. This implies that the correction factor $\gamma$ is allowed to vary across the different individuals. More specifically, consider the conditional hazard model

\begin{eqnarray}\label{GencondHazard}
\tilde{h}_o(t\mid \gamma;\bx) &=& \gamma h_P(A+t;y+t;\bz) + h_E(t;\bx),\\
\gamma &\sim& G.\nonumber
\end{eqnarray}
where $G$ is an arbitrary absolutely continuous cumulative distribution function with support on ${\mathbb R}_+$. The conditional overall survival function is given by
\begin{eqnarray*}
\tilde{S}_o(t\mid\bx,\gamma) &=& \exp\{-\gamma [H_P(A+t;y+t;\bz)-H_P(A;y;\bz)]\}\exp\left[-H_E(t;\bx)\right],\\
\gamma &\sim& G.\nonumber
\end{eqnarray*}
Then, after integrating out the frailty $\gamma$ with respect to the distribution $G$ (see Appendix), the individual marginal overall survival function can be written as
\begin{eqnarray}\label{GeneralSurvival}
\tilde{S}_o(t;\bx) = \exp\{-H_E(t;\bx)\}\Lb_G\{H_P(A+t;y+t;\bz) - H_P(A;y;\bz)\},
\end{eqnarray}
where $\Lb_G\{s\} = \int_0^{\infty} e^{-s  r}dG(r)$ denotes the Laplace transform of $G$ evaluated at time $s$. Next, we consider a specific choice for the distribution $G$: a Gamma distribution. This choice allows for obtaining a closed-form expression of the marginal survival function, in addition to its appealing flexibility and interpretability of parameters.

\subsection*{Gamma Frailty}

Consider the conditional hazard model \eqref{GencondHazard} and suppose that $\gamma \sim \text{Ga}(\mu,b)$, where $\text{Ga}(\mu,b)$ denotes a Gamma distribution with mean parameter $\mu >0$, scale parameter $b>0$, and probability density function $g(r;\mu,b) = \dfrac{r^{\frac{\mu}{b}-1}}{\Gamma\left(\frac{\mu}{b}\right)b^{\frac{\mu}{b}}}\exp\left(-\dfrac{r}{b}\right)$. Then, it follows that
\begin{enumerate}
\item The marginal individual survival function is given by
\begin{eqnarray}\label{OverSurv1}
\tilde{S}_o(t;\bx) &=& \dfrac{\exp\left\{- H_E(t;\bx)\right\} }{\left\{1+b \left[H_P(A+t;y+t;\bz)-H_P(A;y;\bz)\right]\right\}^{\frac{\mu}{b}}}.
\end{eqnarray}
where $H_P(\cdot)$ and $H_E(\cdot)$ represent the cumulative hazards associated to $h_P(\cdot)$ and $h_E(\cdot)$, respectively.
\item The marginal individual hazard function is given by
\begin{eqnarray}\label{OverHazard1}
\tilde{h}_o(t;\bx) =  \dfrac{\mu \, h_P(A+t;y+t;\bz)}{1+b \left[H_P(A+t;y+t;\bz)-H_P(A;y;\bz)\right]} + h_E(t;\bx).
\end{eqnarray}
\end{enumerate}
Expression \eqref{OverHazard1} provides a nice interpretation of our approach since the observed hazard can be seen as a model with a correction function  $\omega_1(t,\bz;\mu,b)=\dfrac{\mu}{1 + b \left[H_P(A+t;y+t;\bz)-H_P(A;y;\bz)\right]}$, which is identifiable and provides a functional form that involves the mean correction, the scale or spread of the correction, and the differences on the population cumulative hazards. Another property of the correction function is that $\lim_{b\rightarrow 0}\omega_1(t,\bz;\mu,b) = \mu$, which indicates that the correction model \eqref{CheuvartHazard} is a limit case. This property, however, does \emph{not} imply that the correction parameter in \eqref{CheuvartHazard} represents the mean correction when $b > 0$, as shown in our simulations. We can also observe that if two patients have the same sociodemographic characteristics (or virtually the same) but different survival times, then the corresponding value of the weight $\omega_1$ will differ since the corresponding differences in the cumulative hazards $H_P(A+t;y+t;\bz)-H_P(A;y;\bz)$ will be different. This, intuitively, suggests that the more extreme the survival time $t$ of a patient is (either too small or too large), compared to other patients with the same sociodemographic characteristics, the more likely this patient has been assigned the incorrect mortality rate from the life table. This approach implicitly assumes that the excess hazard model $h_E(t;\bx)$ is correctly specified and includes all important prognosis factors (see Discussion). It also indicates that the information about the parameters of the frailty distribution is provided by the variability in the observed survival times of similar individuals regarding their sociodemographic characteristics and tumour prognosis factors. This suggests the need for a certain amount of observations for groups of patients with similar characteristics.

\subsection{Hazard structure and baseline hazard}

For models \eqref{HazardAddModel}, \eqref{CheuvartHazard}, and \eqref{OverHazard1}, we adopt the general excess hazard model (GH) proposed in \cite{R18}:
\begin{eqnarray}\label{GH}
h_{\text{E}}^{\text{GH}}\left(t;\bx_i\right) &=& h_0\left(t \exp(\bx_i^{\top}\beta_1)\right)\exp(\bx_i^{\top}\beta_2),\\
H_{\text{E}}^{\text{GH}}\left(t;\bx_i\right) &=& H_0\left(t \exp(\bx_i^{\top}\beta_1)\right)\exp(-\bx_i^{\top}\beta_1+\bx_i^{\top}\beta_2).\nonumber
\end{eqnarray}
where $h_0(\cdot)$ is the baseline hazard. This hazard structure contains, as particular cases, the proportional hazards (PH) model when $\beta_1=0$, the accelerated hazards (AH) model when $\beta_2=0$, the accelerated failure time (AFT) model when $\beta_1=\beta_2$, as well as combinations of these for $\beta_1 \neq \beta_2  \neq 0$. Thus, this structure covers the most popular hazard structures used in the literature, allowing also to capture time-dependent effects through $\beta_1$ (\textit{i.e.}~effects which are not assumed to be constant over the whole follow-up period, such as the effects estimated in PH models). For an extensive discussion on the properties of this GH structure, we refer the reader to \cite{R18}.

The baseline hazard in \eqref{GH} will be modelled using the Exponentiated Weibull (EW) distribution. The Exponentiated Weibull distribution contains three positive parameters $(\kappa,\theta,\alpha)$ (shape, scale, and power), and the corresponding hazard function can capture some basic shapes: increasing, decreasing, unimodal (up-then-down), bathtub (down-then-up) and constant. The Exponentiated Weibull density and cumulative distribution functions with shape, scale, and power parameters $(\kappa,\theta,\alpha)$ are given, respectively, by:
\begin{eqnarray*}
f_{EW}(t) &=&  \alpha \dfrac{\kappa}{\theta} \left(\dfrac{t}{\theta}\right)^{\kappa-1} \left[1-\exp\left\{-\left(\dfrac{t}{\theta}\right)^{\kappa}\right\}\right]^{\alpha-1} \exp\left\{-\left(\dfrac{t}{\theta}\right)^{\kappa}\right\}, \\
F_{EW}(t) &=& \left[1-\exp\left\{-\left(\dfrac{t}{\theta}\right)^{\kappa}\right\}\right]^{\alpha}.
\end{eqnarray*}

Thus, the combination of the GH structure with the choice of the EW baseline hazards can capture a variety of hazard structures, time-dependent effects (through the parameters in $\beta_1$), and baseline hazard shapes, while allowing for a parsimonious implementation of all the proposed models \citep{R18}.

\section{Inference}

\subsection{The classical model}

Let $t_i>0$, $i=1,\dots,n$, be the sample of times to event from a population of cancer patients, with covariates $\bx_i\in {\mathbb R}^p$, and vital status indicators $\delta_i$ ($1$-death, $0$-censored).

For the classical model \eqref{HazardAddModel}, we will rely on the maximum likelihood estimation method, which has been shown to have good inferential properties with the hazard structure detailed in \eqref{GH} \citep{R18}. With this model, the likelihood is defined as:

{\bf M1}. Classical model \eqref{HazardAddModel}:

\begin{eqnarray*}
	{\mathcal L}(\bm{\psi};\text{Data}) &=& \prod_{i=1}^n {h}_o(t_i;\bx_i)^{\delta_i}{S}_o(t_i;\bx_i)\\
	&\propto& \prod_{i=1}^n \left\{ h_P(A_i+t_i; y_i + t_i, \bz_i) + h_E(t_i;\bx_i) \right\}^{\delta_i} \exp\left\{- H_E(t_i;\bx_i)\right\},
\end{eqnarray*}
where $\bm{\psi}$ represent the parameters of the excess hazard model. In our case, $\bm{\psi} = (\kappa, \theta, \alpha, \beta_1, \beta_2)$. Notice that here for model \eqref{HazardAddModel}, we omit the expected survival from the likelihood, as this quantity does not depend on parameters.

\subsection{The models with correction of the background mortality}

For the models of interest, \eqref{CheuvartHazard} and \eqref{OverHazard1}, we will estimate the parameters using the maximum likelihood method. The corresponding likelihood functions are presented below for the single-parameter correction model \eqref{CheuvartHazard} and the frailty correction model \eqref{OverHazard1}

{\bf M2}. Single-parameter correction model \eqref{CheuvartHazard}:
\begin{eqnarray*}\label{LogLikM2}
{\mathcal L}_C(\bm{\psi}_C;\text{Data}) &=& \prod_{i=1}^n \left\{\gamma h_P(A_i+t_i;y_i+t_i;\bz_i) + h_E(t_i;\bx_i) \right\}^{\delta_i} \exp\left\{- H_E(t_i;\bx_i)\right\} \\
&\times& \exp\{-[H_P(A_i+t_i;y_i+t_i;\bz_i)-H_P(A_i;y_i;\bz_i)]\}^{\gamma},
\end{eqnarray*}
where $\bm{\psi}_C = (\gamma, \bm{\psi})$.

{\bf M3}. Frailty correction model \eqref{OverHazard1}:
\begin{eqnarray*}\label{LogLikM3}
{\mathcal L}_F(\bm{\psi}_F;\text{Data}) &=& \prod_{i=1}^n \left\{ \dfrac{\mu \, h_P(A_i+t_i;y_i+t_i;\bz_i)}{1 + b \left[H_P(A_i+t_i;y_i+t_i;\bz_i)-H_P(A_i;y_i;\bz_i)\right]} + h_E(t_i;\bx_i) \right\}^{\delta_i} \\ &\times&\dfrac{\exp\left\{- H_E(t_i;\bx_i)\right\} }{\left\{1 + b \left[H_P(A_i+t_i;y_i+t_i;\bz_i)-H_P(A_i;y_i;\bz_i)\right]\right\}^{\frac{\mu}{b}}},
\end{eqnarray*}
where $\bm{\psi}_F = (\mu,b,\bm{\psi})$.

These likelihood functions can be maximised using standard optimisation routines from the R software (\textit{e.g.}~`nlminb' or `optim'). In order to select between models M1--M3, we use the Akaike Information Criterion (AIC), as these models are estimated using the maximum likelihood method.

\subsection{Choice of initial points for the optimisation process}

For the optimisation process, we consider a two-step algorithm. In the first step, we initialise the search at the  initial values: $\bm{\psi}^{(0)} = (\kappa^{(0)},\theta^{(0)},\alpha^{(0)},\beta_1^{(0)},\beta_2^{(0)})$, where $\kappa^{(0)} = 1$, $\theta^{(0)} = 1$, $\alpha^{(0)} = 2$, $\beta_1^{(0)}=\beta_2^{(0)} = \bm{0}$, and then move from these initial values using one cycle of a coordinate descend algorithm (CDA), coupled with the R command `nlminb' on each step of the CDA (see \citealp{W15} for more details on the CDA). The CDA is an algorithm which successively maximises an objective function along coordinate directions. In our case, we first obtain a new value of the first parameter, $\kappa^{(1)}$, after maximising the objective function with respect to $\kappa$, while setting the other parameter values at their initial values. Then, we maximise again the objective function, obtained by setting this time the initial values to $(\kappa^{(1)},\theta^{(0)},\alpha^{(0)},\beta_1^{(0)},\beta_2^{(0)})$ to get $\theta^{(1)}$. We repeat the same process for all other parameters to finally get a vector $(\kappa^{(1)},\theta^{(1)},\alpha^{(1)},\beta_1^{(1)},\beta_2^{(1)})$. In the second step, we utilise the values obtained with the CDA as new initial values in a general purpose optimisation algorithm (\textit{e.g.}~`nlminb' or `optim' from the R software), in order to obtain the MLE $\widehat{\bm{\psi}}$.

For model M2, we use the initial values $(\gamma,\bm{\hat{\psi}})$, with $\gamma = 1.2$, while for model M3, we employ the initial values $(\mu,b,\bm{\hat{\psi}})$, with $\mu = 1.2$, and $b=0.1$. In practice, we recommend running the optimisation process at several initial points in order to ensure that the global maximum of the likelihood is reached.

\section{Simulation Studies}\label{sec:simulation}

In this section, we present a simulation study where we assess the impact of mismatches in the population hazard on the estimation of the excess hazard, as well as the performance of the proposed correction models (M2-M3). The true values of the parameters are chosen in order to produce scenarios that resemble cancer population studies concerning an aggressive type of cancer (relatively low 5-year net survival, see Appendix Table 1), such as lung cancer, similar to the simulations presented in \cite{R18}. An additional extensive simulation scenario resembling a less aggressive type of cancer (such as colon cancer, see Appendix Table 2) is presented in the Appendix (Simulation design II).

\subsection{Data Generation}
Briefly, in the Simulation Design I, we simulated $N=1000$ data sets of size $n= 5000, 10000$, assuming the additive hazard decomposition given in \eqref{HazardAddModel}. The variable ``age'' was simulated as a continuous variable using a mixture of uniform distributions with $0.25$ probability on $(30,65)$, $0.35$ probability on $(65,75)$, and $0.40$ probability on $(75,85)$ years old. The binary variables ``sex'' and ``W'' were both simulated from a binomial distribution with probability $0.5$ (the binary variable ``W'' could be viewed as ``treatment'', or ``comorbidity'', or ``stage'' (early and late)). In all scenarios, we simulated the ``other-causes'' time to event using the UK life tables based on ``age'' and ``sex'' (assuming that all patients were diagnosed on the same year). The time to event from the excess hazard (cancer death time) was generated using the inverse transform method, assuming effects of the 3 variables ``age'', ``sex''  and ``W'' and an Exponentiated Weibull distribution. We assumed either (i) only administrative censoring at $T_C=5$ years, which induced approximately $25\%$ censoring in all cases, or (ii) an additional independent random censoring (drop-out) using an exponential distribution with rate parameter $r$, inducing approximately $30\%$ censoring in these cases. Given the GH structure \eqref{GH} adopted for the simulation, all variables affect the time scale (\textit{i.e.}~time-dependent effects) as well as the hazard scale (\textit{i.e.}~changing the level of the hazard). We refer the reader to \cite{R18} for a more detailed discussion on the interpretation of the GH structure.

\begin{figure}[]
\begin{center}
\begin{tabular}{c}
\includegraphics[scale=0.5]{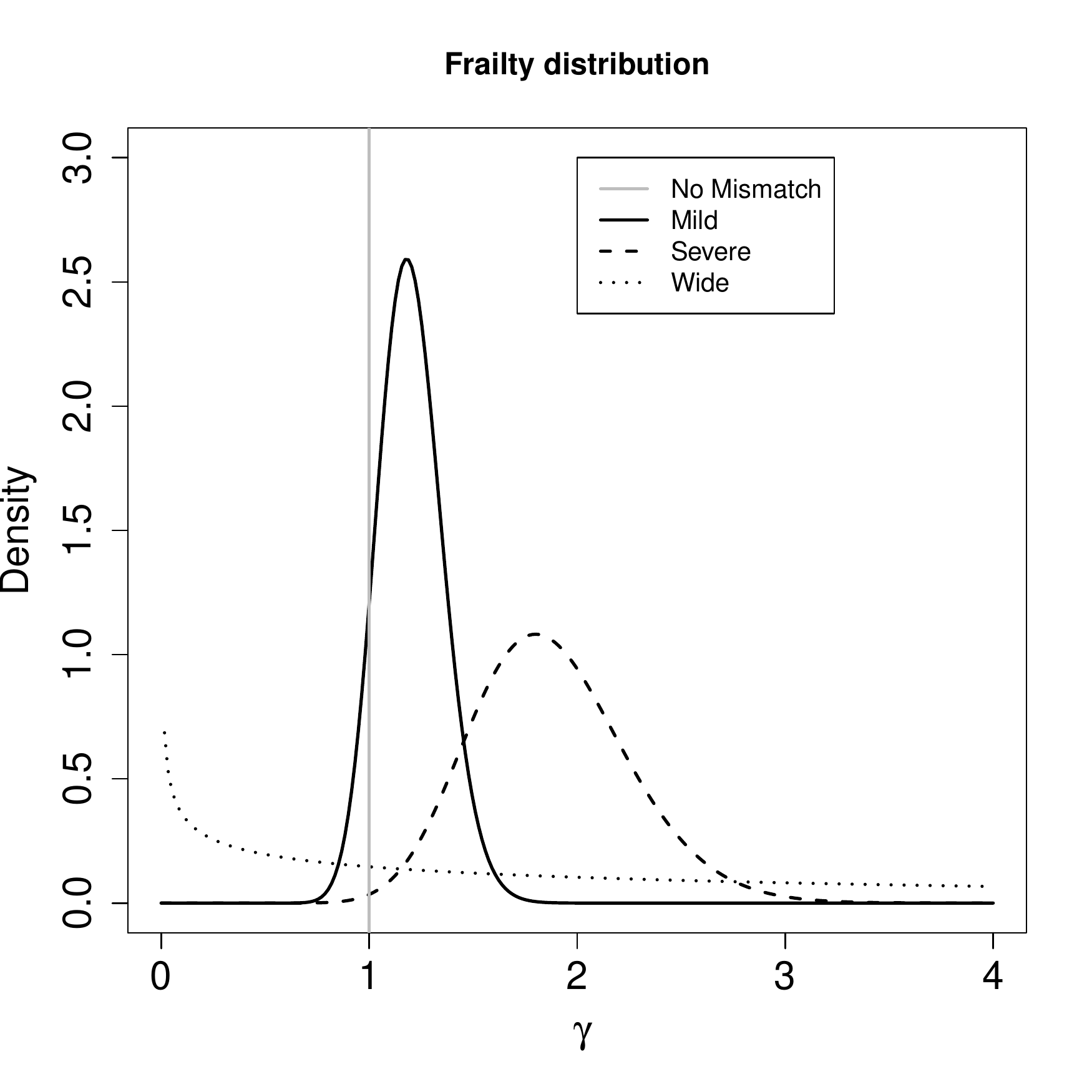}
\end{tabular}
\end{center}
\caption{Frailty distributions used in the simulation.}
\label{fig:frailties}
\end{figure}

We consider 4 scenarios that represent mismatches in the life tables: (i) No mismatch, where the data are generated from model \eqref{HazardAddModel}; (ii) Moderate mismatch, where the data are generated from the hazard model \eqref{GencondHazard} with $\gamma \sim \text{Ga}(1.2,0.02)$; (iii) Severe mismatch, where the data are generated from the hazard model \eqref{GencondHazard} with $\gamma \sim \text{Ga}(1.875,0.075)$; and (iv) Wide mismatch, where the data are generated from the hazard model \eqref{GencondHazard} with $\gamma \sim \text{Ga}(6.5,10)$ (see Figure \ref{fig:frailties}). In all of these scenarios we fit models M1--M3. We also consider selecting between models M1--M3 using AIC, in order to identify the model favoured by the data. The estimates of the parameters of the excess hazard model selected using AIC are reported, and this model is referred to as model M4. For model M4, we report the estimated correction parameter $c$ associated to the selected model. This is, $\widehat{c}=1$ if AIC selects model M1, $\widehat{c}=\widehat{\gamma}$ if AIC selects model M2, and $\widehat{c}=\widehat{\mu}$ if AIC selects model M3. We report coverage of the asymptotic normal confidence intervals in all the scenarios.

\subsection{Simulation Results}
\begin{figure}[!htbp]
\begin{center}
\textbf{Design I: $\gamma\sim Ga(6.5,10)$}
\end{center}
\begin{center}
\begin{tabular}{c}
\includegraphics[scale=0.85]{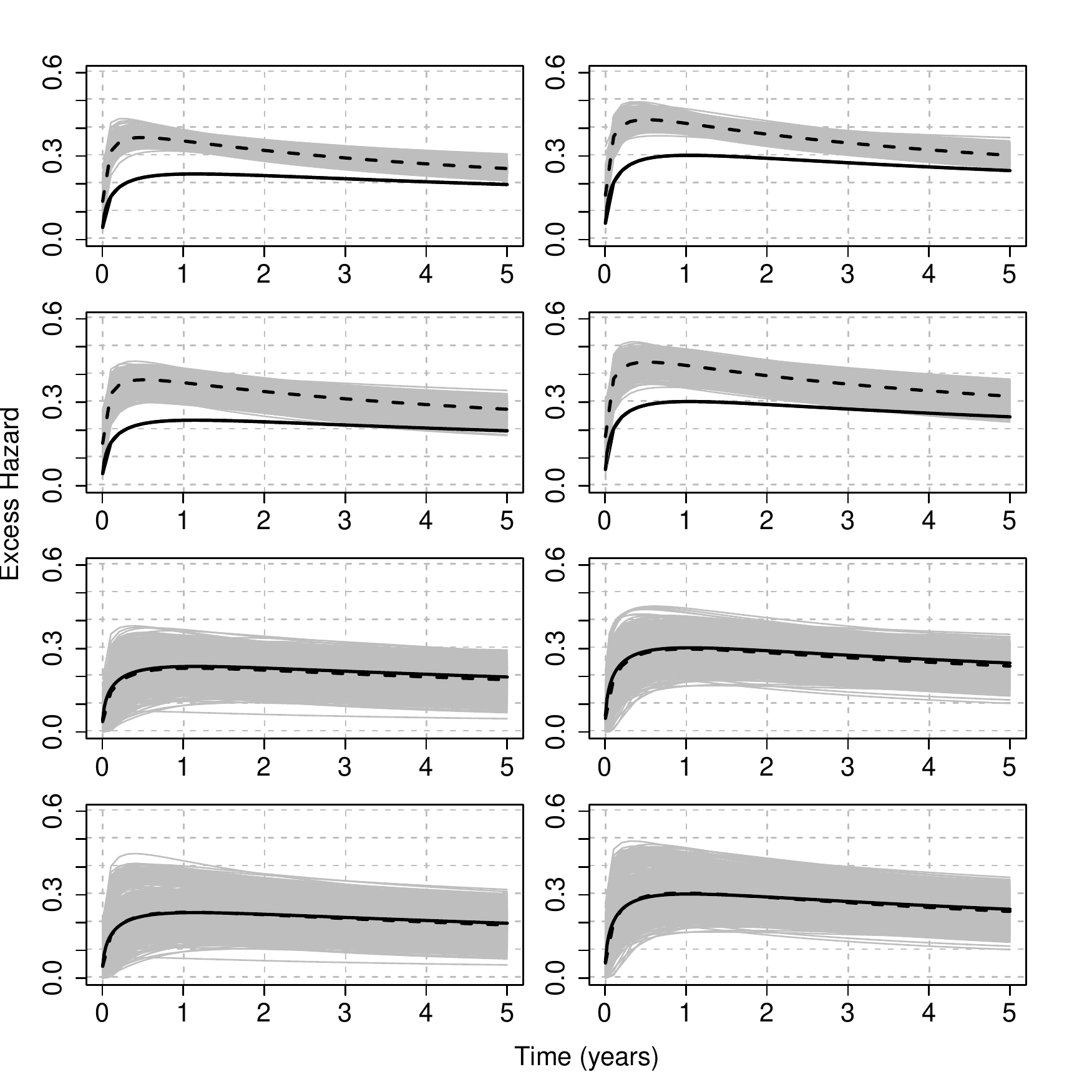}
\end{tabular}
\end{center}
\caption{Scenario with wide mismatch: $\gamma\sim Ga(6.5,10)$. Models M1--M4 from top to bottom. Mean of the fitted excess hazards (dashed lines), compared to the true generating excess hazard (continuous lines), and 1000 sample-specific fitted excess hazards (grey lines) for $n=5000$ and $30\%$ censoring. Panels from left to right correspond to two sets of values for the covariates (age, sex, comorbidity)=$(70,0,0),(70,0,1)$, respectively.}
\label{fig:Realistic5000_1}
\end{figure}

For illustration, Table \ref{table:Realistic5000_1} and Figure \ref{fig:Realistic5000_1} show the results for the scenario with $n=5000$, 30\% of censoring and a Wide mismatch (iv). The results of the remaining scenarios are shown in the Appendix, as well as the results with sample size of $n=10000$ and the simulation design II with a higher net survival. As expected, we observe that when there is Severe and Wide mismatch (iii)--(iv), the model without correction (M1) will lead to biased parameters estimates, as well as poor coverage (Table \ref{table:Realistic5000_1}, Figure \ref{fig:Realistic5000_1}, Appendix Tables 5, 8, 9 and Appendix Figures 4, 7, 8). This is reflected in the MLEs of the model parameters as well as on the fitted excess hazards. The bias of the parameter estimates and the poor coverages of M1 are more pronounced in the simulation design II with high net survival (Appendix Tables 12, 13, 16, 17 and Appendix Figures 11, 12, 15, 16).
In scenarios (i) and (ii) with No or Moderate mismatch, the fitted correction models M2 and M3 are centred around the true generating model, although they exhibit (as expected) a slightly larger variability compared to model M1 (Appendix Tables 3, 4, 6, 7, 10, 11, 14, 15, and Appendix Figures 2, 3, 5, 6, 9, 10, 13, 14).
In scenario (iii) with Severe mismatch, the fitted correction models M2 and M3 properly correct the mismatch as these models are centred around the true generating model, with the cost of higher variability. The parameters estimated with model M1 are biased with a very low coverage (Appendix Tables 5, 8, 12, 16 and Appendix Figures 4, 7, 11, 15).
In scenario (iv) with Wide mismatch, the fitted models M1 and M2 are biased and far from the true generating model. On the other hand, model M3 can capture this mismatch and reduce the bias properly (Table \ref{table:Realistic5000_1}, Figure \ref{fig:Realistic5000_1}, Appendix Tables 9, 13, 17 and  Appendix Figures 8, 12, 16).
The models selected with AIC (M4) are also centred around the true generating model in all scenarios. In cases with no mismatch or moderate mismatch, we observe that the bias on the estimates is small, the coverage is close to $95\%$, and all models tend to be relatively close to the true generating model. In those cases, the AIC tends to favour model M1. Table 18 in the Appendix shows the proportion of selected models using AIC. Overall, we can see that M3 is favoured in scenario (iv) with Wide mismatch. In the scenario (iii) with Severe mismatch, the proportion of selected models depends on the contribution of the excess hazard compared to the population hazard. In general, selecting the models using AIC or another model selection tool is advised in order to identify the need for correcting the population hazard. The simulation study results obtained with $N=10000$ per sample were very similar to the ones obtained with $N=5000$.

\begin{table}[!htbp]
\begin{center}
\textbf{Design I: $\gamma\sim Ga(6.5,10)$}
\end{center}
\footnotesize
\centering
\begin{tabular}{|cccccccc|}
\hline
Model & Parameter & MMLE & mMLE & ESD & Mean Std Error & RMSE & Coverage \\
\hline
\multirow{ 9}{*}{M1} & $\sigma$ (1.75) & 1.212 & 1.214 & 0.228 & 0.218 & 0.584 & 0.464 \\
&  $\kappa$ (0.6) & 0.592 & 0.593 & 0.046 & 0.045 & 0.047 & 0.934 \\
&  $\alpha$ (2.5) & 2.365 & 2.315 & 0.328 & 0.308 & 0.355 & 0.887 \\
&  $\beta_{11}$ (0.1) & 0.119 & 0.118 & 0.012 & 0.012 & 0.023 & 0.700 \\
&  $\beta_{12}$ (0.1) & 0.409 & 0.405 & 0.214 & 0.220 & 0.376 & 0.711 \\
&  $\beta_{13}$ (0.1) & -0.034 & -0.030 & 0.233 & 0.221 & 0.268 & 0.902 \\
&  $\beta_{21}$ (0.05) & 0.065 & 0.065 & 0.002 & 0.002 & 0.015 & 0.000 \\
&  $\beta_{22}$ (0.2) & 0.296 & 0.296 & 0.047 & 0.045 & 0.107 & 0.449 \\
&  $\beta_{23}$ (0.25) & 0.161 & 0.159 & 0.047 & 0.046 & 0.101 & 0.484 \\
\hline
\multirow{ 10}{*}{M2} & $\sigma$ (1.75) & 1.284 & 1.308 & 0.232 & 0.243 & 0.520 & 0.741 \\
&  $\kappa$ (0.6) & 0.625 & 0.635 & 0.060 & 0.065 & 0.065 & 0.845 \\
&  $\alpha$ (2.5) & 2.199 & 2.106 & 0.412 & 0.402 & 0.510 & 0.689 \\
&  $\beta_{11}$ (0.1) & 0.119 & 0.118 & 0.012 & 0.013 & 0.023 & 0.715 \\
&  $\beta_{12}$ (0.1) & 0.411 & 0.406 & 0.215 & 0.223 & 0.378 & 0.717 \\
&  $\beta_{13}$ (0.1) & -0.011 & -0.001 & 0.225 & 0.225 & 0.251 & 0.928 \\
&  $\beta_{21}$ (0.05) & 0.066 & 0.067 & 0.003 & 0.003 & 0.017 & 0.004 \\
&  $\beta_{22}$ (0.2) & 0.308 & 0.309 & 0.046 & 0.047 & 0.117 & 0.352 \\
&  $\beta_{23}$ (0.25) & 0.155 & 0.152 & 0.044 & 0.045 & 0.105 & 0.412 \\
&  $\gamma$ (6.5) & 0.453 & 0.001 & 0.652 & 0.549 & 6.082 & 0.846 \\
\hline
\multirow{ 11}{*}{M3} &  $\sigma$ (1.75) & 1.406 & 1.323 & 0.745 & 0.721 & 0.820 & 0.978 \\
&  $\kappa$ (0.6) & 0.543 & 0.546 & 0.127 & 0.122 & 0.139 & 0.966 \\
&  $\alpha$ (2.5) & 4.082 & 2.992 & 7.019 & 3.070 & 7.192 & 0.970 \\
&  $\beta_{11}$ (0.1) & 0.099 & 0.099 & 0.028 & 0.024 & 0.028 & 0.913 \\
&  $\beta_{12}$ (0.1) & 0.121 & 0.134 & 0.340 & 0.339 & 0.340 & 0.959 \\
&  $\beta_{13}$ (0.1) & 0.089 & 0.103 & 0.354 & 0.326 & 0.354 & 0.950 \\
&  $\beta_{21}$ (0.05) & 0.048 & 0.049 & 0.009 & 0.009 & 0.010 & 0.948 \\
&  $\beta_{22}$ (0.2) & 0.187 & 0.198 & 0.089 & 0.089 & 0.090 & 0.954 \\
&  $\beta_{23}$ (0.25) & 0.262 & 0.251 & 0.087 & 0.085 & 0.087 & 0.949 \\
&  $b$ (10) & 13.060 & 9.072 & 12.247 & 10.379 & 12.618 & 0.847 \\
&  $\mu$ (6.5) & 7.010 & 7.134 & 1.839 & 1.727 & 1.908 & 0.853 \\
\hline
\multirow{ 10}{*}{M4} &  $\sigma$ (1.75) & 1.367 & 1.274 & 0.733 & 0.686 & 0.826 & 0.920 \\
&  $\kappa$ (0.6) & 0.540 & 0.546 & 0.122 & 0.117 & 0.136 & 0.968 \\
&  $\alpha$ (2.5) & 3.992 & 2.978 & 6.479 & 2.694 & 6.645 & 0.960 \\
&  $\beta_{11}$ (0.1) & 0.100 & 0.100 & 0.028 & 0.023 & 0.028 & 0.895 \\
&  $\beta_{12}$ (0.1) & 0.139 & 0.151 & 0.341 & 0.328 & 0.343 & 0.938 \\
&  $\beta_{13}$ (0.1) & 0.082 & 0.092 & 0.350 & 0.317 & 0.350 & 0.945 \\
&  $\beta_{21}$ (0.05) & 0.049 & 0.049 & 0.010 & 0.008 & 0.010 & 0.859 \\
&  $\beta_{22}$ (0.2) & 0.194 & 0.203 & 0.092 & 0.086 & 0.093 & 0.904 \\
&  $\beta_{23}$ (0.25) & 0.257 & 0.248 & 0.089 & 0.082 & 0.089 & 0.907 \\
&  $c$ (6.5) & 6.709 & 7.137 & 2.470 & -- & 2.477 & -- \\
\hline
\end{tabular}
\caption{\footnotesize  Simulation results for the scenario GH with $(\sigma,\kappa,\alpha) = (1.75,0.6,2.5)$, $\beta_1 = (0.1,0.1,0.1)$, $\beta_2 = (0.05,0.2,0.25)$, $n=5000$, and wide mismatch $\gamma \sim Ga(6.5,10)$. Mean of the MLEs (MMLE), median of the MLEs (mMLE), empirical standard deviation (ESD), mean (estimated) standard error, root-mean-square error (RMSE), and coverage proportions (Coverage).}
\label{table:Realistic5000_1}
\end{table}

\pagebreak
\newpage
\section{Application: Lung Cancer data}\label{sec:Application}

We now analyse a dataset obtained from population-based national cancer registry of Non-Small Cell Lung Cancer (NSCLC) patients diagnosed in 2012 in England. For deriving information on stage at diagnosis and presence of comorbidities at the time of diagnosis, we linked these data to administrative data (Hospital Episode Statistics -HES- and the Lung Cancer Audit data -LUCADA-) and then applied specific algorithms  \citep{benitez2016, maringe2017}. We used a 6-year period up to 6 months before diagnosis to retrieve information on comorbidity. We checked for the presence of cardiovascular comorbidity (at least one of: Myocardial infarction, Congestive heart failure, Peripheral vascular disease, and Cerebrovascular disease) and Chronic Obstructive Pulmonary Disease (COPD). We measured deprivation using the Income Domain from the 2010 England Indices of Multiple Deprivation, defined at the Lower Super Output Area level (mean population $1500$). The Income Domain measures the proportion of the population in an area experiencing deprivation related to low income, and ranges from $1\%$ to $75\%$ in our data (\url{https://www.gov.uk/government/statistics/english-indices-of-deprivation-2010}). Follow-up was assessed on the 31st of December 2015, at which time patients alive were censored (so the maximum follow-up was 4 years). We restricted our analysis to men with no missing data. We observed $n=15688$ patients with complete cases among which $n_o=13603$ died before the 31st of December 2015, and 17 patients were lost to follow-up (censored before the 31st of December 2015). The median follow-up among patients censored was $3.45$ years, mainly because of administrative censoring. The $25\%$, $50\%$ and $75\%$ quantiles of the patients' age at diagnosis was $65.8$, $73.0$, $80.0$ while the mean was $72.5$. Among the patients, $2210 $ were diagnosed at Stage I, $1502$ at Stage II, $3679$ at Stage III, and $8297$ at Stage IV. Finally, $3224$ patients were classified with a cardiovascular comorbidity and $3154$ with a chronic obstructive pulmonary disease.

\begin{figure}[]
	\begin{center}
		\begin{tabular}{c}
			\includegraphics[scale=0.75]{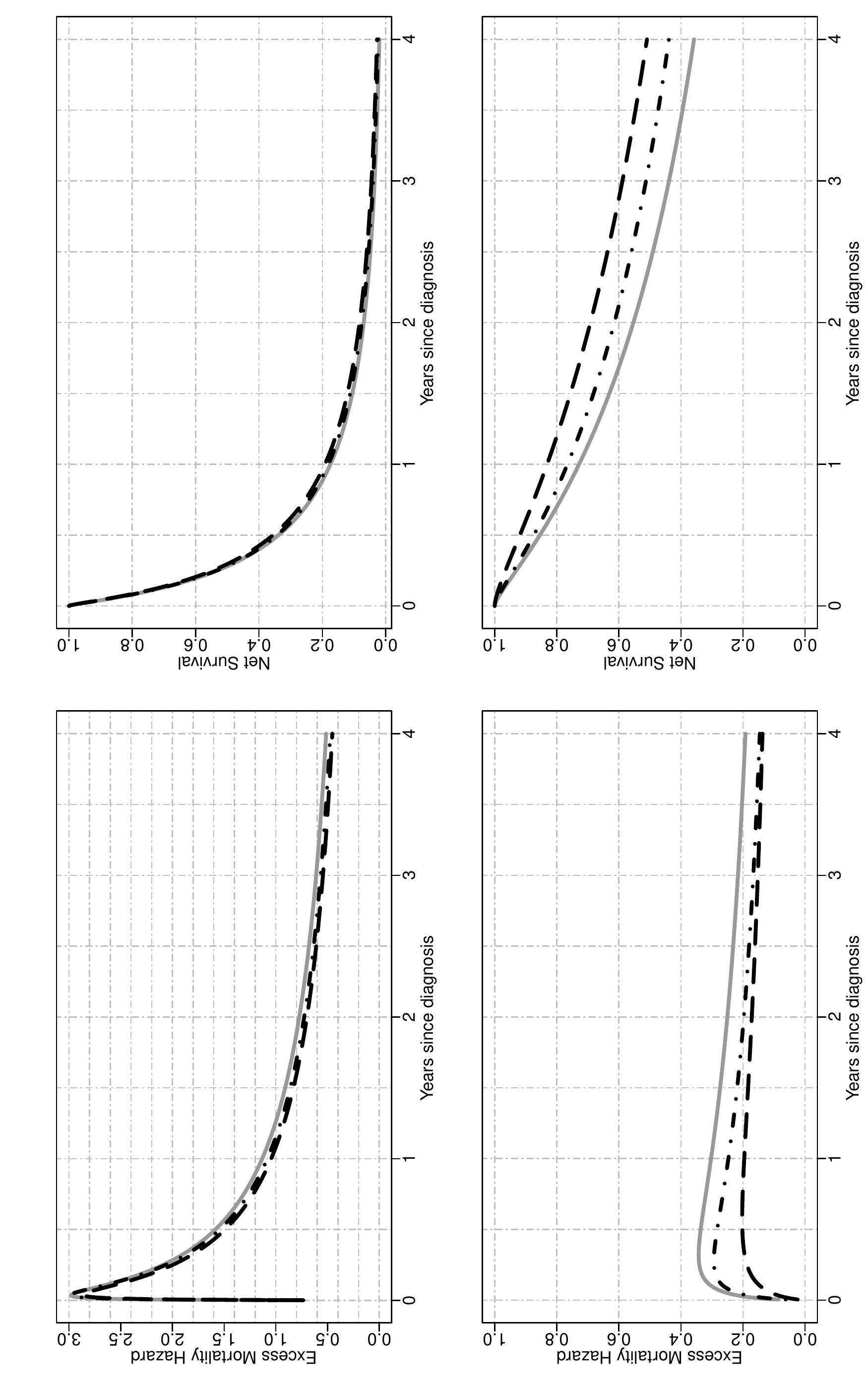}
		\end{tabular}
	\end{center}
	\caption{Illustration for lung cancer patients: Excess mortality hazard (left panels) and the corresponding net survival (right panels) for men aged 70 years at diagnosis, with Income Deprivation score equals to 0.1 (\textit{i.e.}~least deprived), without Cardiovascular comorbidity nor COPD, and with stage IV cancer at diagnosis (upper panels) or stage II cancer at diagnosis (lower panels). M1=solid grey lines, M2=dot-dashed black lines, M3=long-dashed black lines}
	\label{fig:EMH_NS}
\end{figure}

We applied models M1--M3 to estimate the excess mortality hazard using deprivation-specific life tables (detailed by sex, age, year and Government Office Region in addition to the deprivation quintile). Consequently, we are implicitly assuming that the variables age, sex, deprivation, tumour stage, and comorbidity accurately explain the excess hazard in NSCLC patients. The regression parameter estimates for the excess hazard models M1--M3, as well as the correction parameters (for models M2--M3), are reported in Table \ref{table:betahat}. For illustrating the results, the excess mortality hazard and the corresponding Net Survival for two pre-defined subgroups of patients are depicted in Figure \ref{fig:EMH_NS}.

In Table \ref{table:betahat}, between the three models (M1, M2 and M3), the AIC favours model M3 (\textit{i.e.}~the frailty correction model). Differences on $\beta$ estimates (regression coefficients) between M1 and M3 are substantial. For example the protective effect of Stage I cancer (compared to being diagnosed with a Stage IV cancer) is even higher when accounting for mismatched life tables. This interpretation follows by noticing that the two parameters associated with Stage I are negative (see \citealp{R18} for details on those hazard-structure models and their interpretation). The impact of the presence of a comorbidity is higher in M3 compared to M1 and M2. Thus, correcting the population life table for unobserved predicting variables of background mortality seems to be quite relevant in this example. An unobserved variable which certainly affects the population mortality hazard here is smoking status. We observe that the frailty distribution, used for correcting the population mortality in M3, cumulates $23\%$ of the probability mass below $1$, and $77\%$  above $1$. That is, the value $1$ represents the $23\%$ quantile of the fitted Gamma frailty distribution with scale parameter $9.83$ and mean $6.54$. These values are in fact related to the proportion of smokers (roughly $80\%$, which would, in principle, require a correction higher than 1) for England lung cancer patients \citep{ellis2014}, which provides an intuitive interpretation of the frailty distribution parameters. This interpretation, of course, has to be taken only at an intuitive level since the correction induced with the frailty model M3 is not interpretable in terms of a single missing characteristic, but it represents a combination of missing characteristics such as drug use (most likely tobacco in this case), the presence of comorbidities among other lifestyle related diseases and its impact on the general population mortality, and etcetera.

In order to evaluate the effect of not having deprivation-specific life tables, we have also fitted excess hazard models using life tables without the deprivation variable (\textit{i.e.}~national life tables). The results obtained with the deprivation-specific life tables and those obtained with the national life tables (excluding deprivation) are very similar (see section 5, Table 19, in the Appendix), which is due to the high lethality of lung cancer (inducing negligible differences of population mortality between deprivation groups). Figure 1 in the Appendix shows the net survival curves obtained for the whole population with models M1--M3 as well as the non-parametric Pohar-Perme estimator \citep{PS12}. We observe that model M1 and the Pohar-Perme estimator are virtually the same, which indicates that M1 can properly capture time-dependent effects \citep{R18}, an assumption made for the correction models. Model M2 produces a net survival curve which is consistently above that obtained with M1. The net survival curve obtained with M3 is above all others, which is explained by the fact that most of the probability mass ($77\%$ ) of the frailty correction is above $1$. We compared estimates of the excess mortality hazard and the corresponding net survival for two subgroups of patients (Figure \ref{fig:EMH_NS}). For stage-IV patients, the differences between each model is almost not visible (upper panels), while the difference between models M1--M3 could be more clearly seen in stage II patients subgroup (lower panels).

\begin{table}[]
	\centering
	\begin{tabular}{|rlll|}
		\hline
		& M1 & M2 & M3  \\
		\hline
  $b$ & -- & -- & 9.83 (3.03)  \\
  $\gamma \mid \mu$ & -- & 2.7 (0.21) & 6.54 (0.91) \\
  $\theta$ & 0.05 (0.01) & 0.03 (0.01) & 0.03 (0.01)  \\
  $\kappa$ & 0.38 (0.01) & 0.35 (0.01) & 0.34 (0.01)  \\
  $\alpha$ & 4.64 (0.34) & 5.64 (0.48) & 5.92 (0.58)  \\
  Age-t & 0.29 (0.04) & 0.29 (0.04) & 0.16 (0.05)  \\
  Dep-t & 0.11 (0.04) & 0.12 (0.04) & 0.09 (0.04)  \\
  Stage 1-t & -2.66 (0.25) & -2.17 (0.32) & -5.4 (1.4)  \\
  Stage 2-t & -2.2 (0.2) & -2 (0.22) & -2.69 (0.35)  \\
  Stage 3-t & -1.66 (0.11) & -1.57 (0.11) & -1.75 (0.13)  \\
  CV-t & 0.31 (0.11) & 0.31 (0.11) & 0.42 (0.11) \\
  COPD-t & 0.13 (0.11) & 0.08 (0.12) & 0.37 (0.14)  \\
  Age & 0.27 (0.01) & 0.23 (0.02) & 0.16 (0.02)  \\
  Dep & 0.06 (0.01) & 0.06 (0.01) & 0.04 (0.01)  \\
  Stage 1 & -2.84 (0.06) & -3.13 (0.1) & -3.53 (0.36)  \\
  Stage 2 & -2.16 (0.06) & -2.32 (0.07) & -2.65 (0.1)  \\
  Stage 3 & -1.23 (0.03) & -1.27 (0.04) & -1.36 (0.04)  \\
  CV & 0.24 (0.04) & 0.26 (0.04) & 0.3 (0.04)  \\
  COPD & 0.19 (0.04) & 0.17 (0.04) & 0.25 (0.05)  \\
  AIC & 20304.69 & 20241.27 & {\bf 20213.41} \\
  \hline
	\end{tabular}
\caption{\small Regression parameter estimates (standard errors) using models M1--M3, with their corresponding AIC on the men lung cancer dataset. Note: The time- dependent effects are indicated with ‘-t’. For model M2, $\gamma$ is estimated, while $\mu$ is estimated for model M3. Age=Age at diagnosis (centred at 70, and divided by 10), Dep=Income Deprivation Score (centred at 0.1, and divided by 10), CV=CardioVascular comorbidity, COPD=Chronic Obstructive Pulmonary Disease, AIC=Akaike Information Criteria (best model indicated in bold font).}
\label{table:betahat}
\end{table}

\section{Discussion}
\subsection{Summary of findings}
Using the general hazard structure in \cite{R18}, we have proposed excess hazard regression models that can account for mismatches in the life tables induced by the unavailability of information on relevant population characteristics. The correction models based on a frailty distribution account for non-specific mismatches in the life table, in the sense that the correction is not associated to the lack of known specific variables for constructing the life tables, but to the effect of potentially several unavailable characteristics, which is allowed to be different for each patient. This is the main difference with Cheuvart's model \eqref{CheuvartHazard}, which we used here for comparison purpose even though this model was mainly developed in the context of a randomised clinical trial when a selection bias of the patient population is expected. Thus, Cheuvart's model assumes the same constant correction parameter for all patients, which is the main difference with our proposed frailty correction models used in population-based cancer registry data. We have shown that the proposed frailty correction models are able to properly identify and correct these mismatches in several simulation scenarios, provided that the sample size is large enough ($5000$ or more). Not accounting for mismatched life tables in the relative survival setting may lead to inappropriate net survival comparisons between populations. The need for relatively large samples in order to identify mismatches in the life tables is unsurprising, as there may be several reasons why general population life tables are not fit for our cancer patient population (drug use, lifestyle related diseases, deprivation, ethnicity, and etcetera), and only a large enough sample would guarantee that the data contains enough information to adjust for the unavailability of those variables. Intuitively, the information about these correction parameters is provided by the sample of differences of the population cumulative hazards $ H_P(A_i+t_i;\bz_i) - H_P(A_i;\bz_i)$. The implicit assumptions behind the proposed correction model M3 \eqref{OverHazard1} are:
\begin{enumerate}[(i)]
	\item The set of covariates $\bx$ includes the relevant cancer-specific variables. Thus, all missing information (if any) is produced from a mismatch of the population mortality rate.
	\item The model $h_E(t;\bx)$ is properly specified. This is, the fitted excess hazard model is flexible enough to approximate the excess hazard.
\end{enumerate}
Assumption (i) reflects the fact that the model was constructed to only capture mismatches in the life tables, since the correction parameter only affects the population mortality hazard. For instance, in our lung cancer data example, we assume that the excess hazard is accurately explained by age, sex, deprivation, tumour stage, and comorbidity. However, a potential risk factor that is not included there is the smoking status, which is not available at the population level in England. This risk factor might affect both the background mortality hazard (because of diseases or complications associated with smoking) and the cancer-related mortality hazard, in the case that patients continue smoking after the diagnosis of cancer. Indeed, smoking is a driver of lung cancer incidence, but its impact on cancer-related mortality hazard may not be that clear because comorbidity conditions already accounts (at least partially) to smoking-related complications. Thus, the main interest of including smoking status in the predictor variables of the cancer-related mortality hazard would be for patients who continue smoking after the diagnosis of cancer, and it would certainly be interesting to explore the effect of including this variable in the model once it becomes available. Assumption (ii) is important since model misspecification can also affect the correction made on the population hazard. If a covariate which appears in $\bz$ and $\bx$ is wrongly modelled in $h_E(t;\bx)$ (\textit{e.g.}~not accounting for time-dependent effects), this may also affect the correction. In principle, this is not an onerous condition since one would usually aim at properly modelling the excess hazard, which is typically the main function of interest. Moreover, recent developments in the use of splines and parametric models \citep{R02,G03,C07,R07, Charvat16,remontet:2018,R18} allow for a tractable inclusion of nonlinear and time-dependent effects in excess hazard models. We also assume that there is enough heterogeneity about the unobserved variables of interest. For instance, if we want to assess the impact of mismatched life tables in terms of deprivation, we assume that the sample contains large enough numbers of individuals with different deprivation levels. The amount of data required to accurately estimate the parameters of the correction models has been explored through a simulation study. Certainly, we would not recommend trying to correct for mismatches in the life tables in samples containing substantially fewer than $5000$ observations, or with high censoring rates (\textit{e.g.} higher than $50\%$). Overall, we have found that model M3 is a good option for accounting for mismatches in the life tables, provided a large enough sample. Its use however is not automatic, and should be analysed on a case by case basis. Comparing the results between corrected (M3) and uncorrected (M1) models, as well as the nonparametric Pohar-Perme estimator, is advisable in practice, in addition to the use of expert knowledge from clinicians or epidemiologists in order to understand and explain the source of the mismatches.

\subsection{Other models in the literature: shared and correlated frailty models}
\cite{Z97} describes two extensions of long-term excess hazards models, where the main goal is to account for an increased risk of dying of other diseases in patients with certain cancers.  The first extension consists of a shared frailty model in which a random effect (frailty) is multiplied by the population hazard and the excess hazard as follows:
\begin{eqnarray}\label{SFM}
h_o^{S}(t;\bx\mid \gamma) &=& [h_P(A+t;y+t,\bz) + h_E(t;\bx)]{\gamma},
\end{eqnarray}
where $\gamma\sim G$, and $G$ is a distribution with positive support, typically chosen to be a Gamma distribution with unknown shape and scale parameters. Perhaps unsurprisingly, the induced model is non-identifiable unless the random effect has unit mean. Given that the Gamma distribution is asymmetric, the assumption of unit mean implies that ${\mathbb P}(\gamma\leq 1) > {\mathbb P}(\gamma >1)$, which may not be a reasonable assumption in some scenarios since this implies that there is a higher probability of requiring a shrinking correction to the population hazard ($\gamma \leq 1$) than an increasing one ($\gamma>1$). For a general framework of frailty hazard models we refer the reader to \cite{A08}.

\cite{Z97} proposed a correlated frailty model, by using frailties on both the population hazard and the excess hazard:
\begin{eqnarray}\label{CFM}
h_o^{Z}(t;\bx\mid \gamma_1,\gamma_2) &=& h_P(A+t;y+t,\bz){\gamma_1} + h_E(t;\bx){\gamma_2},
\end{eqnarray}
where $(\gamma_1,\gamma_2)\sim G_2$, and $G_2$ is a bivariate distribution with support on the positive quadrant. Intuitively, it is difficult (if at all feasible) to obtain information about the factors affecting the population hazard and the excess hazard (which is typically a flexible parametric model), and the dependencies between them, simultaneously. In fact, \cite{Z97} found that the maximum likelihood estimators of the parameters of model \eqref{CFM} do not exist, suggestive of identifiability issues of this model. \cite{Z97} proposed a number of restrictions of the parameter space (to a compact set) in order to alleviate these estimation issues. However, even after those restrictions, the MLE was on the boundary of the restricted parameter space, which suggests remaining lack of identifiability.

\subsection{Further research}

From the results of our simulation study, we have observed a larger variability in the estimators of additional parameters corresponding to the correction of the background mortality hazard. In order to reduce this variability, penalised maximum likelihood estimation methods could be used to shrink the correction parameter (\emph{i.e.}~for M2 or for M3) towards the value $1$. This will be explored in future research.

From the simulation study, the coverage proportions of the additional parameter correcting the life table were lower than the nominal value. Using a robust estimator of the variance for this additional parameter may be an option to reach a better coverage, as may be calculating profile likelihood intervals.

Another extension of model \eqref{CheuvartHazard} consists of modelling the correction parameters $\gamma$ and $\mu$ in terms of a set of covariates, say ${\bf w}$. A related approach has recently been studied in \cite{touraine:2018}. Possible limitations include the inferential challenges in estimating $q \geq 2$ (the dimension of ${\bf w}$) when the sample size is not large enough. In addition, the assumption of proportional population hazards is often too restrictive in the cancer survival field. In practice, one natural question is whether the Gamma frailty distribution is flexible enough to model the random correction. Using maximum likelihood estimation implies that the estimators of the parameters of the frailty distribution will converge to the values that minimise the distance (in fact, the Kullback-Leibler divergence) to the true generating model. Section 6, Appendix Tables 20-21 and Appendix Figures 17-18, shows a simulated example where the random correction is simulated from a lognormal distribution (instead of Gamma). This example indicates that model M3 has a good performance even if the random corrections are not generated from a Gamma distribution, but as long as the Gamma distribution can approximate the shape of the true generating distribution. A possible extension consists of using a more flexible frailty distribution with a tractable Laplace transform, in order to obtain tractable expressions for the hazard and cumulative hazard functions. An attractive option is the power variance function (PVF) family of distributions \citep{A08}, which contains three parameters instead of two. This, of course, complicates the estimation process.

\section*{Software}

Software in the form of R code, together with a sample input data set and complete documentation is available under request, and an R Markdown document entitled ``Simulation design I: Excess hazard models for insufficiently stratified life tables'' is available on the website \url{http://www.rpubs.com/FJRubio/FGH} and the GitHub repository \url{https://github.com/FJRubio67/ExcessHazardModels}.

\section*{Acknowledgements}

The authors thank two referees, an Associate Editor, and the Editor for helpful comments.

We obtained the ethical and statutory approvals required for this research (PIAG 1-05(c)/2007; ECC 1-05(a)/2010); ethical
approval updated 6 April 2017 (REC 13/LO/0610). We attest that we have obtained appropriate permissions and paid any
required fees for use of copyright protected materials.

\bibliographystyle{plainnat}
\bibliography{references}

\begin{thebibliography}{24}
\providecommand{\natexlab}[1]{#1}
\providecommand{\url}[1]{\texttt{#1}}
\expandafter\ifx\csname urlstyle\endcsname\relax
  \providecommand{\doi}[1]{doi: #1}\else
  \providecommand{\doi}{doi: \begingroup \urlstyle{rm}\Url}\fi

\bibitem[Aalen et~al.(2008)Aalen, Borgan, and Gjessing]{A08}
O.~Aalen, O.~Borgan, and H.~Gjessing.
\newblock \emph{Survival and event history analysis: a process point of view}.
\newblock Springer-Verlag, New York, 2008.

\bibitem[Benitez-Majano et~al.(2016)Benitez-Majano, Fowler, Maringe,
  Di~Girolamo, and Rachet]{benitez2016}
S.~Benitez-Majano, H.~Fowler, C.~Maringe, C.~Di~Girolamo, and B.~Rachet.
\newblock Deriving stage at diagnosis from multiple population-based sources:
  colorectal and lung cancer in {E}ngland.
\newblock \emph{British Journal of Cancer}, 115:\penalty0 391, 2016.

\bibitem[Bower et~al.(2018)Bower, Andersson, Crowther, Dickman, Lambe, and
  Lambert]{bower:2018}
H.~Bower, T.M.L. Andersson, M.J. Crowther, P.W. Dickman, M.~Lambe, and P.C.
  Lambert.
\newblock Adjusting expected mortality rates using information from a control
  population: An example using socioeconomic status.
\newblock \emph{American Journal of Epidemiology}, 187:\penalty0 828--836,
  2018.

\bibitem[Charvat et~al.(2016)Charvat, Remontet, Bossard, Roche, Dejardin,
  Rachet, Launoy, and Belot]{Charvat16}
H.~Charvat, L.~Remontet, N.~Bossard, L.~Roche, O.~Dejardin, B.~Rachet,
  G.~Launoy, and A.~Belot.
\newblock A multilevel excess hazard model to estimate net survival on
  hierarchical data allowing for non-linear and non-proportional effects of
  covariates.
\newblock \emph{Statistics in Medicine}, 35:\penalty0 3066--3084, 2016.

\bibitem[Cheuvart and Ryan(1991)]{CR91}
B.~Cheuvart and L.~Ryan.
\newblock Adjusting for age-related competing mortality in long-term cancer
  clinical trials.
\newblock \emph{Statistics in Medicine}, 10:\penalty0 65--77, 1991.

\bibitem[Dickman et~al.(1998)Dickman, Auvinen, Voutilainen, and Hakulinen]{D98}
P.W. Dickman, A.~Auvinen, E.T. Voutilainen, and T.~Hakulinen.
\newblock Measuring social class differences in cancer patient survival: is it
  necessary to control for social class differences in general population
  mortality? a {F}innish population-based study.
\newblock \emph{Journal of Epidemiology and Community Health}, 52:\penalty0
  727--734, 1998.

\bibitem[Ellis et~al.(2014)Ellis, Coleman, and Rachet]{ellis2014}
L.~Ellis, M.P. Coleman, and B.~Rachet.
\newblock The impact of life tables adjusted for smoking on the socio-economic
  difference in net survival for laryngeal and lung cancer.
\newblock \emph{British Journal of Cancer}, 111:\penalty0 195, 2014.

\bibitem[Esteve et~al.(1990)Esteve, Benhamou, Croasdale, and Raymond]{E90}
J.~Esteve, E.~Benhamou, M.~Croasdale, and L.~Raymond.
\newblock Relative survival and the estimation of net survival: elements for
  further discussion.
\newblock \emph{Statistics in Medicine}, 9:\penalty0 529--538, 1990.

\bibitem[Giorgi et~al.(2003)Giorgi, Abrahamowicz, Quantin, Bolard, Est\`eve,
  Gouvernet, and Faivre]{G03}
R.~Giorgi, M.~Abrahamowicz, C.~Quantin, P.~Bolard, J.~Est\`eve, J.~Gouvernet,
  and J.~Faivre.
\newblock {{A} relative survival regression model using {B}-spline functions to
  model non-proportional hazards}.
\newblock \emph{Statistics in Medicine}, 22:\penalty0 2767--2784, 2003.

\bibitem[Graff{\'e}o et~al.(2012)Graff{\'e}o, Jooste, and Giorgi]{G12}
N.~Graff{\'e}o, V.~Jooste, and R.~Giorgi.
\newblock The impact of additional life-table variables on excess mortality
  estimates.
\newblock \emph{Statistics in Medicine}, 31:\penalty0 4219--4230, 2012.

\bibitem[Maringe et~al.(2017)Maringe, Fowler, Rachet, and
  Luque-Fernandez]{maringe2017}
C.~Maringe, H.~Fowler, B.~Rachet, and M.A. Luque-Fernandez.
\newblock Reproducibility, reliability and validity of population-based
  administrative health data for the assessment of cancer non-related
  comorbidities.
\newblock \emph{PloS One}, 12:\penalty0 e0172814, 2017.

\bibitem[Nelson et~al.(2007)Nelson, Lambert, Squire, and Jones]{C07}
C.P. Nelson, P.C. Lambert, I.B. Squire, and D.R. Jones.
\newblock Flexible parametric models for relative survival with application in
  coronary heart disease.
\newblock \emph{Statistics in Medicine}, 26:\penalty0 5486--5498, 2007.

\bibitem[Pavli{\v{c}} and Pohar-Perme(2018)]{P18}
K.~Pavli{\v{c}} and M.~Pohar-Perme.
\newblock Using pseudo-observations for estimation in relative survival.
\newblock \emph{Biostatistics}, in press:\penalty0 na--na, 2018.

\bibitem[Perme et~al.(2012)Perme, Stare, and Est{\`e}ve]{PS12}
M.P. Perme, J.~Stare, and J.~Est{\`e}ve.
\newblock On estimation in relative survival.
\newblock \emph{Biometrics}, 68:\penalty0 113--120, 2012.

\bibitem[Pohar~Perme et~al.(2009)Pohar~Perme, Henderson, and Stare]{Pohar2009}
M.~Pohar~Perme, R.~Henderson, and J.~Stare.
\newblock {{A}n approach to estimation in relative survival regression}.
\newblock \emph{Biostatistics}, 10:\penalty0 136--146, 2009.

\bibitem[Pohar-Perme et~al.(2016)Pohar-Perme, Est{\`e}ve, and
  Rachet]{perme:2016}
M.~Pohar-Perme, J.~Est{\`e}ve, and B.~Rachet.
\newblock Analysing population-based cancer survival--settling the
  controversies.
\newblock \emph{BMC Cancer}, 16:\penalty0 933, 2016.

\bibitem[Remontet et~al.(2007)Remontet, Bossard, Belot, and Esteve]{R07}
L.~Remontet, N.~Bossard, A.~Belot, and J.~Esteve.
\newblock An overall strategy based on regression models to estimate relative
  survival and model the effects of prognostic factors in cancer survival
  studies.
\newblock \emph{Statistics in Medicine}, 26:\penalty0 2214--2228, 2007.

\bibitem[Remontet et~al.(2018)Remontet, Uhry, Bossard, Iwaz, Belot, Danieli,
  Charvat, and Roche]{remontet:2018}
L.~Remontet, Z.~Uhry, N.~Bossard, J.~Iwaz, A.~Belot, C.~Danieli, H.~Charvat,
  and L.~Roche.
\newblock Flexible and structured survival model for a simultaneous estimation
  of non-linear and non-proportional effects and complex interactions between
  continuous variables: Performance of this multidimensional penalized spline
  approach in net survival trend analysis.
\newblock \emph{Statistical Methods in Medical Research}, in press:\penalty0
  na--na, 2018.

\bibitem[Royston and Parmar(2002)]{R02}
P.~Royston and M.K.B. Parmar.
\newblock Flexible parametric proportional-hazards and proportional-odds models
  for censored survival data, with application to prognostic modelling and
  estimation of treatment effects.
\newblock \emph{Statistics in Medicine}, 21:\penalty0 2175--2197, 2002.

\bibitem[Rubio et~al.(2018)Rubio, Remontet, Jewell, and Belot]{R18}
F.J. Rubio, L.~Remontet, N.P. Jewell, and A.~Belot.
\newblock On a general structure for hazard-based regression models: an
  application to population-based cancer research.
\newblock \emph{Statistical Methods in Medical Research}, in press:\penalty0
  na--na, 2018.

\bibitem[Touraine et~al.(2019)Touraine, Graff{\'e}o, Giorgi, and the CENSUR
  working~survival group]{touraine:2018}
C.~Touraine, N.~Graff{\'e}o, R.~Giorgi, and the CENSUR working~survival group.
\newblock More accurate cancer-related excess mortality through correcting
  background mortality for extra variables.
\newblock \emph{Statistical Methods in Medical Research}, in press:\penalty0
  na--na, 2019.

\bibitem[Woods et~al.(2005)Woods, Rachet, Riga, Stone, Shah, and Coleman]{W05}
L.M. Woods, B.~Rachet, M.~Riga, N.~Stone, A.~Shah, and M.P. Coleman.
\newblock Geographical variation in life expectancy at birth in england and
  wales is largely explained by deprivation.
\newblock \emph{Journal of Epidemiology and Community Health}, 59:\penalty0
  115--120, 2005.

\bibitem[Wright(2015)]{W15}
S.J. Wright.
\newblock Coordinate descent algorithms.
\newblock \emph{Mathematical Programming}, 151:\penalty0 3--34, 2015.

\bibitem[Zahl(1997)]{Z97}
P.H. Zahl.
\newblock Frailty modelling for the excess hazard.
\newblock \emph{Statistics in Medicine}, 16:\penalty0 1573--1585, 1997.

\end{thebibliography}

\end{document}